# Scripting Relational Database Engine Using Transducer

Feng Tian
VitesseData Inc.

ftian@vitessedata.com

## ABSTRACT

We allow database user to script a parallel relational database engine with a procedural language. Procedural language code is executed as a user defined relational query operator called transducer. Transducer is tightly integrated with relation engine, including query optimizer, query executor and can be executed in parallel like other query operators. With transducer, we can efficiently execute queries that are very difficult to express in SQL. As example, we show how to run time series and graph queries, etc, within a parallel relational database.

## 1 INTRODUCTION

Analytic workload in the "Big Data" era is getting more and more complex. Many new applications demand deep analytics driven by breakthrough in AI or new data models such as graph. Many people believe that these new workloads are not good fit for a relational database because,

1. The data is not relational (tabular).
2. The query is not easy to express in SQL and/or the result SQL cannot be efficiently executed.

The first point itself is an interesting research topic that is out of the scope of this paper. Here we just want to point out that firstly, there is large amount of data stored in relational database and these data is often of high value. Secondly, some people's "unstructured data" can be highly structured, tabular data in other's view. For example, web log data is about who clicked which link and large amount of IoT data is about which sensor recorded some event, when, where. Sometimes, "semi-structured" data format like JSON can be treated as a user defined data type, often used to encode sparse tuples.

This paper focuses on the second point. SQL, and its theoretic foundation relational algebra/calculus, can express queries of first order logic. With extensions like OLAP Window functions and recursive query processing, SQL has achieved Turing Completeness, but this has little value in practice because it is extremely hard to write such SQL queries, for example, to express an algorithm querying graph data.

Imagine we need to run a shortest path query on the graph of a user profile database of a social network. Currently there are a few options,

1. Use a special purpose graph database. Such a special purpose database usually is confined in a relatively small market segment. Because first order logic is so fundamental, eventually such databases will ship an embedded (often immature) relational query engine.
2. Build a graph processor on top of relational database. Graph algorithms are mapped to relational operators and executed by relational engine (maybe using temporary tables for staging results). This approach is often inflexible, and inefficient.
3. Process data in database, dump data out, run graph algorithm with another tool, then load data back into database for further processing. Besides obvious performance issues, user often need to handle data refresh and version mismatch.
4. Use cursors and stored procedure. Unfortunately, cursor kills performance – it is a slow row by row interface and often forcing a parallel database to bring large amount of data onto one node.

We implemented a new way to extend relational database engine capability called Transducer. Transducer is a query operator just like other relational operators such as join, except it is implemented by user using a procedural language. As part of execution plan, transducer cooperates with query optimizer and can be scheduled parallelly at run time.

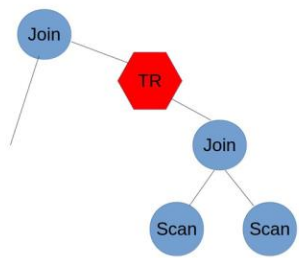

**Figure 1: Transducer (TR node) in a query plan.**

It is worth point out that Transducer is different from a user defined function (UDF) in that while user can implement UDF in procedural language, UDF is usually invoked on a per tuple base (or per group base if it is a user defined aggregate function). Transducer is a query operator and sees all input tuples. It can carry an execution context with full knowledge of the history of data it has seen. This execution context can be cleanly implemented using local variables on the stack of procedural language. The ability to maintain such a context across tuple or group boundary is crucial and will be clear in the later use case section.

In the rest of the paper, Section 2 will describe the syntax and semantics of transducer. We will also describe the implementation of transducer in Deepgreen, a parallel relational database. Section 3 describes several real-world use cases. Section 4 reviews related work. We summarize in Section 5.

## 2 SYNTAX, SEMANTICS AND IMPLEMENTATION OF TRANSDUCER IN DEEPGREEN

First, we briefly introduce the Deepgreen database and some terminologies related to the parallel execution architecture. Deepgreen is a massively parallel processing (MPP) rational database system based on the open source Greenplum database [14]. Inspired by [6] and [8], Deepgreen implemented a JIT query engine using LLVM and achieved 2-10 times speedup across all the queries in TPC-H benchmark. The overall architecture of Deepgreen is a classical shared-nothing parallel database that is very similar to GAMMA [5]. Each Deepgreen database has a *master*, which accepts user connections, parses and optimizes SQL queries. Data is stored in *segment*s and query is executed in parallel on all *segment*s. During query processing, when necessary, data is exchanged between segments (or gathered to one segment if the semantics of the query requires) via network. Deepgreen tries to spread data of a table and the intermediate results of a query evenly across all segments using a *distribution policy*, usually by hashing some columns.

### 2.1 Syntax

User can write transducer as a user defined function (UDF) in the select clause of a SQL query, but the similarity between transducer and ordinary UDF stops at SQL parser. When SQL parser detects the transducer function, instead of adding an expression that compute values of the UDF, a transducer operator is inserted into the query plan as illustrated in Figure 1.

As an example, Program Listing 1 shows a convoluted way of writing a query that is equivalent to "select id, txt from t where id % 3 = 1". We will walk through this example in details to let reader get familiar with the syntax. We will only show sketches of program listing in later examples.

① declares the output columns of the transducer. The argument to transducer_col_type function specifies the ordinal number of the

```
Select transducer_col_int4(1) as id,      ①
       transducer_col_text(2) as txt,
       transducer($$PHIExec go            ②
// The following is a valid go program    ③
// BEGIN INPUT
// id int32
// t text
// END INPUT
// BEGIN OUTPUT
// id int32
// t text
// END OUTPUT
package main
func main() {
  for rec := NextInput(); rec != nil; rec =
NextInput() {                             ④
    id, _ := rec.Get_id()                 ⑤
    t, _ := rec.Get_t()
    if id % 3 == 1 {                      ⑥
        var outrec OutRecord
        outrec.Set_id(id)
        outrec.Set_t(t)
        WriteOutput(&outrec)              ⑦
    }
    WriteOutput(nil)                      ⑧
  }
} $$),
t.id, t.txt                               ⑨
From t                                    ⑩
     Program Listing 1: Simple Transducer.
```

transducer output field that user wants to project. We use different functions for different return types because SQL is a static, strong typed language. Projected result can be wrapped in a subquery for further processing.

② defines the transducer. $$ starts a SQL multiline raw string. Deepgreen supports user defined transducer as covered in this paper as well as some system built-in transducers. Built-in transducers are implemented using C and user defined transducer can be implemented using python (a dynamic typed interpreted language) or go (a static typed compiled language). PHIExec (PHI stands for Pretty High Integration) will invoke a user defined transducer, in this example, is implemented using Go.

③ till the end of the raw string is a valid Go program. At the beginning of the go program is a comment section for input output types. Deepgreen will generate code at run time using these type directives. Both input and output struct types are generated for Go program.

④ is a for loop to read all input. NextInput function is generated at run time with proper input record type.

⑤ reads the fields of input record. _ sign is Go syntax for a place holder for unused variable, in this case, whether the field is null.

⑥ is the filter, id % 3 = 1.

⑦ writes the output record. WriteOutput is also generated to use correct struct type.

⑧ writes nil to indicate end of output.

⑨ expressions in the select clause after the transducer function are treated as input columns to the transducer function.

⑩ The from clause is the input table to transducer. It can be a table or a subquery.

There are quite some lines of boilerplate code on input output typing but they are straightforward and could be automatically filled by an editor tools like intellisense of an IDE. In this simple example, we call NextInput and WriteOutput in a loop, but they can be called in arbitrary order. For example, transducer can output data before getting first input record and can output many records after exhausting all input records.

The UDF syntax is chosen so that syntactically a transducer query is valid SQL. User can wrap the transducer in subquery or view just like any other SQL queries. All database clients like JDBC, ODBC can still work without any compatibility issue.

## 2.2 Optimizer and Execution Semantics

When SQL parser detects the transducer function, it inserts a transducer operator into plan tree. Transducer serves as an optimizer barrier, that is, the input subquery to transducer and the further processing of transducer output are optimized independently. Transducer receives stats and sorting order from its input subquery and can use this information to optimize its own code. Transducer in turn will supply stats and sorting order back to optimizer for later query optimization.

After query optimization, transducer is an operator of a physical query plan and is dispatched to proper segments for execution. Deepgreen may run a slice of a query plan parallelly on all segments, or gather all data to one segment, or run on master. Transducer operators are scheduled in the same way as other relational operators.

For user defined transducers, Deepgreen first run a code generation step to generate proper input and output data structure, then an optional compilation step if the procedural language is a compiled language. User defined transducers are executed in a forked process and communicate with query execution engine via pipes. Input and output records are batched as row groups. Built-in transducers are implemented in C and executed within the database engine.

In general, a transducer has many instances running in parallel on all segments. Just as other relational operator, transducer will see all the input data but there is no guarantee on which segment an input record will be processed and there is no guarantee of the tuple arriving order. However, user can enforce a data distribution policy between segments and a sorting order using SQL partition by and order by clause. Enforcing such constraints let transducer leverage highly optimized sorting algorithm of a database system and we will illustrate with an example in next Section.

Because the transducer is just an operator, diagnostic tools like explain (display query plan) and explain analyze (displays query plan with execution statistics) all work as expected. This is very important in practice for user to analyzing and fixing query performance problems.

```
                        QUERY PLAN
--------------------------------------------------------------
 Gather Motion 2:1  (slice1; segments:
2)  (cost=0.00..2.05 rows=2 width=4)
   -> Transducer  (cost=0.00..2.05 rows=1 width=4)
       ->  Seq Scan on eachseg  (cost=0.00..2.05 rows=1 width=4)
 Optimizer status: legacy query optimizer
(4 rows)
 Program Listing 2: Explain output of a query
                    with transducer.
```

Program listing 2 shows an explain output of the simple query in Program listing 1. Transducer is run in parallel on all segments and gather motion will gather transducer output to master then return to user. The statistics such as cost, estimated number of rows are correctly displayed.

## 3  USE CASES

In this section, we will demonstrate how transducer can be used in real world. All example code in this section can be downloaded from https://github.com/vitesse-ftian/dgnb. Deepgreen database software can be downloaded from our website http://www.vitessedata.com/ and also available on AWS Market Place.

### 3.1 Access External Data Source and Data Migration

Transducer in Deepgreen can be implemented in python or Go, therefore, it can read any external data source as long as the data source has a python or Go driver. For example, transducer can access other databases like Oracle, or NoSQL databases like

MangoDB, Redis, elastic search or CSV files, or streaming data using Kafka. Depending on the capability of the external data source provider, Deepgreen can run the transducer parallelly on each segment, or, just using one on master.

Deepgreen provides another technology (XDrive) to access external data source. To use XDrive, user can mount the external data source as a "drive", then create an external table in database to read or write the "drive". XDrive has a plugin system that also allows user to code external data accessor using any programming language. External table requires user to write DDL for external data source and install precompiled plugin binaries. Compared to XDrive or other similar external table based methods, transducer is quite verbose due to boilerplate code. However, transducer offer extradentary flexibility. In customer sites, with a few lines of code, we can gain access to data sources that we never encountered before.

As a special example, we implemented the data transfer and migration tool between two Deepgreen clusters using a built-in transducer. The tool is extremely simple, basically for each data transfer task we execute two queries; one transducer on source database that "send" data and one on target database that "recv" data. The tool is very flexible in that the data source can be a table or a query and the target can do some translation before insert into a table. In our performance test, we transferred 1TB TPC-H data in 10 minutes (roughly 10 million tuples per second) using 4 machines each with 2 10GigE network cards. Transducer saturated network bandwidth with only moderate CPU load.

### 3.2 Generate Runs

Suppose you have a database of stock prices *Stock(symbol, date, price)* and you want to find all stocks with a gaining streak of 10 days or more, or the longest gaining or losing streak of each stock. This query is very hard to express in SQL but straightforward if using transducer, as in Program Listing 3. This example introduced several interesting concepts and worth a closer look at the code.

① uses a WITH clause to put the transducer in a subquery so later processing is easier to write. Below that, are input output type boilerplate code.

② declares a local variable on stack to keep track of current run. This is the context of the transducer. The context keeps a history of the input we have seen so far.

③ assumes inputs to transducer are partitioned by symbol and within each symbol the inputs are ordered by trading date.

④ starts a run, by creating a new outrec and saving state.

⑤ determines if it is still in the same run by comparing price with the information recorded in outrec, the context.

⑥ if the run continues, only need to update context. Otherwise output record and start a new run.

⑦ is the most important concept illustrated in this query. We used an OLAP window specification to enforce data is partitioned by symbol, therefore all data for each individual stock will be processed at the same segments. Within each partition the data is ordered by date. Partition and order by are executed by well-tuned database hash/sort operators.

⑧ further processing from subquery declared by WITH clause. It can be any SQL, for example, aggregate like MAX/AVG of (endprice – beginprice).

This example demonstrated two key difference between transducer and an ordinary user defined function.
1. Transducer can keep and maintaining the context of history of inputs,
2. Transducer can choose when to generate an output (or many outputs), depending on the state of its context.

Another important technique illustrated in this example is to specify data partition and ordering in the SQL query. Such partition and ordering are enforced by database query optimizer and the partition and order information is carried into transducer, or, from another point of view, transducer can leverage the partition and sorting functionality of a relational database.

### 3.3 Graph Query with BSP

Bulk Synchronous Parallel (BSP) [9] is a model for designing parallel algorithms. A BSP algorithm has a set of computing processes run concurrently. These processes can exchange data via network. The parallel algorithm is broken into super steps and at the end of a super step all processes participate in a barrier synchronization. Google Pregel [7] is a graph analytic engine using BSP. BSP is also used by open source graph analytic engines like Apache Giraph [11].

Deepgreen transducer supports BSP using the following set of APIs,
- *BspInit* initializes the BSP system.
- *BspSend* sends data to peers. A process can send data to itself.
- *BspNext* reads data sent by peers
- *BspSync* is the barriers synchronization. At the end of each super step all processes must call this method. This method will vote if the algorithm is complete. BSP algorithm stops when all processes voted complete in this super step.

Just like the NextInput and WriteOutput function, transducer does proper code generation for data structures and manages batching row to row groups in BspSend and BspNext.

```
WITH run AS ( select                           ①
Transducer_col_text(1) as symbol,
Transducer_col_int4(2) as begin,
Transducer_col_float8(3) as beginprice,
Transducer_col_int4(3) as end,
Transducer_col_float8(4) as endprice,
Transducer($$PHIExec go
// … input and output types …
var outrec *OutRecord                          ②
for r:=NextInput(); r!=nil; r=NextInput() {
  symbol, _ := r.Get_symbol()
  day, _ := r.Get_day()
  price, _ := r.Get_price()
  if day == 0 {                                ③
    // new symbol, output prev run and start
    // a new run
    if outrec != nil {
       WriteOutput(outrec)
    }
    outrec = new(OutRecord)                    ④
    outrec.Set_symbol(symbol)
    outrec.Set_begin(day)
    outrec.Set_beginprice(price)
    outrec.Set_end(day)
    outrec.Set_endprice(price)
  } else {
    // is still a run?                         ⑤
    Isuprun := …
    Isdownrun := …
    If isuprun || isdownrun {
       Outrec.Set_end(day)                     ⑥
       Outrec.Set_endprice(price)
    } else {
       WriteOutput(rec)
       … // start a new run
    }
  }
}
if outrec != nil {
  WriteOutput(outrec)
}
WriteOutput(nil)
…
$$), t.symbol, t.day, t.price FROM
( select row_number() over
  (partition by symbol order by day),          ⑦
  Symbol, day, price
  FROM stock
) t)
SELECT stock from run where end – day > 10     ⑧
```

**Program Listing 3: Compute Stock Runs**

```
… boilerplate type code …
… suppose graph edge is (i, j)
func main() {
  BspInit(2)                                   ①

  // SuperStep 1, redistribute                 ②
  For r:=NextInput(); r!=nil; r=NextInput() {
    BspSend(i%2, r)
    BspSend(j%2, r)
  }
  BspSync(false)

  // SuperStep 2, build graph.  Graph is       ③
  // a map from node id to out edges.
  // Each node has a depth, initialized to -1
  // except start node of the search (depth 0)
  BspSend(start%2, start)
  BspSync(false)

  For Sstep := 2; ; sstep++ {                  ④
   For rr:=BspNext(); rr!=nil; rr=BspNext() {
     // BSF, if depth is -1, mark it
     // BspSend to next SuperStep
   }
   // If no node is marked in this step, vote
   // true, otherwise vote false
   Done := BspSync(ifNoMarkThisStep)
   If Done {
     Break
   }
  }

  For node in graph {                          ⑤
    WriteOutput(id, depth)
  }
  WriteOutput(nil)
}
… boilerplate code
FROM Graph;
```

**Program Listing 4. BFS Using BSP**

Program List 4 contains a BSP example of breadth first search (BFS) on a graph. BFS is a simple graph algorithm that can be found in any algorithm book. The pseudo code listing mainly shows how BSP code is used to parallelize the execution.

① init BSP system with 2 processes.

② the first super step just read input edges and redistribute to a BSP process according to a hash function (mod). In this step, BspSync voted that it has more work to do.

③ builds the graph in memory as hash table. Do proper initialization on depth (-1 to indicate a node that has not been visited by the search).

④ is the BSP algorithm which follows edges and marks unvisited nodes with correct depth. We loop over super steps. Each iteration deepens search depth. If a super step does not mark any node, BspSync votes no more works, otherwise it votes to have more work. Note that in a super step if a process voted no more work, but some of its peers voted to have more work, BspSync will return as not finished. The process needs to loop back to next step.

⑤ outputs node with its depth visited during BSF. If a node is not connected with start node, its depth will be -1.

We can further process the BSF result, for example, count nodes in each depth level. We use the DBLP co-authorship graph [13] as an example. The graph is an undirected graph. Nodes are authors and two nodes are connected when the they co-author a paper. The graph has more than 300 thousand nodes and more than 1 million edges. The BFS on this graph can be executed on a mid-range laptop in just 2 or 3 seconds. After counting nodes at each depth, we get a chart in Figure 2.

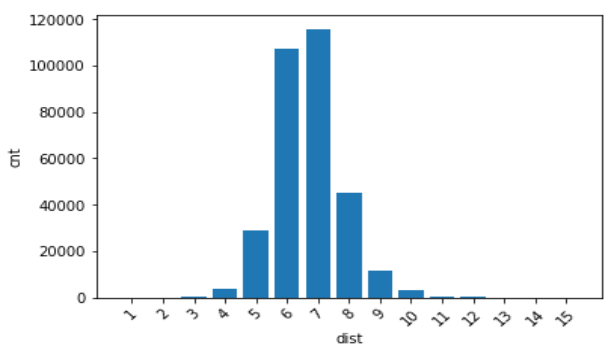

**Figure 2. DBLP Co-Authorship Graph Depth**

We noticed that the node id of authors has some gaps. We assume that is because some authors just wrote a paper without collaborating with a peer. However, it is interesting to observe that the result of BSF does not have any node at depth -1, that means, once a research collaborates with another people, he is connected to the whole community!

Single source shortest path (SSSP) is another important query on graph data. Dijkstra's algorithm is hard to parallelize, but the Bellman-Ford algorithm is well suited for execution using BSP. The implementation of Bellman-Ford algorithm is almost the same as BFS – differences are only a few lines. When looping over super steps, instead of marking the depth of breadth first search, Bellman-Ford algorithm simply relaxes distance on all the edges. We skip the code listing for this example and reader can find the code on our website.

SQL supports recursive query processing using WITH clause and Bill of Materials (BOM) is the most used example. BOM can be considered as a special case of the traversal algorithm on graph – traversal starts from root of a tree structure. Recursive query like BOM is relatively hard to implement efficiently in a parallel database, but straightforward with transducer. Another difficult problem for recursive query is to detect the stop condition of the recursion. Many database manuals have a "user beware" footnote on recursive query. Detecting stop condition is much more natural with a procedural language. For example, BSF handles cycle in graph with marking a flag. It is easy to add a check for negative weighted cycles to the Bellman-Ford algorithm.

### 3.4  Deep Learning with TensorFlow

We have seen recent breakthrough in AI and Deep learning and our users want to bring these technology into database. TensorFlow [1] from Google is one of the leading deep learning tools. [16] contains an extremely well written tutorial on Tensorflow. The tutorial illustrates how to train a regression model or neural network to predicate a point is blue or green. We ported the example in the tutorial to Deepgreen using transducer. One query trains a neural network with three hidden layers and the other one uses the trained model to do inference on data.

We made only a few changes to the example in the tutorial,
1. Instead of having three sets of data ('linear', 'moon', 'saturn'), we mixed all the data in one table.
2. We parallelized the code.
3. The old code read/parse a CSV file to read input, we replaced that code with transducer NextInput() call.

We did not change the "interesting" part of neural network model itself. Python code is simply copy/pasted into transducer. The interesting idea related to transducer is quite similar in training and inference, therefore, Program listing 5 only shows the SQL to do inference.

① puts the TensorFlow predication result in a subquery. The predication column is the predicated color by TensorFlow, tag is the real color of the point.

② uses python, which is the most widely used language binding for TensorFlow. Go binding for TensorFlow is available but not as mature.

③ reads input data into batches.

④ the original tutorial has three data sets and each data sets is trained, evaluated separately. We combine the three data sets in one table and turn them into categorical attributes. This part uses SQL case expression to turn a categorical attribute to a floating-point value to be used by the neural network.

⑤ further processes predication result with SQL, for example, here we want to examine the points that is predicated incorrectly.

```
WITH EVAL AS (                              ①
Select transducer_col_int4(1) as prediction,
   Transducer_col_int4(2) as tag,
   Transducer_col_float4(3) as x,
   Transducer_col_flaot4(4) as y,
   Transducer($$PHIExec python            ②
… boilerplate code …

Def nextbatch(batch):                       ③
  While true:
   If cnt == BATCH_SIZE:
      Break
   Rec = NextInput()
   If not rec:
      Break
   Cnt += 1
   batch.append(rec)

… same tensorflow code …
… use nextbatch() to get input data …
… use WriteOutput() …
… to feed data backed to database …
$$), data.*
FROM (
   Select tag,
          Case when cat = 'linear' then 1.0  ④
               When cat = 'moon' then 2.0
               When cat = 'saturn' then 3.0
          End,
          X, y from points
) data
)
Select * from EVAL where predication <> tag  ⑤
```

**Program Listing 5. Use TensorFlow to do prediction**

We want to put emphasis on step ③, that is, turning inputs into batches. Batching is critical for performance when using TensorFlow and other deep learning tools. Batching is relatively simple in transducer (our code is actually easier and shorter than the code parsing CSV file in original example) but surprisingly hard to do in SQL without transducer. Without transducer, the "obvious" way of using TensorFlow is either call a python UDF row by row, or, aggregate all data in an aggregate UDF. They are either slow, or can cause pressure on resource like memory.

### 3.5 Quick Prototype Tool

Finally, transducer is a very handy tool for R&D prototype. When a developer has a new join algorithm, or storage format, or index, he can quickly implement a transducer and test it inside a MPP database. For example, Deepgreen has a very efficient, scalable sampling algorithm. Transducer allows us to quickly implement a proof of concept and validate the design. Later we transferred the implementation into database engine and added SQL language support.

## 4  RELATED WORK

Most relational databases support user defined functions, aggregate functions and table functions. For example, Greenplum or PostgreSQL PL/Language supports many popular programming languages. UDF in general is not as flexible, or as powerful, as user defined query operator.

Relational databases have some mechanisms to access external data source, such as external tables, dblink, OLEDB. These mechanisms usually implement a table scan operator at the leaf of plan tree and cannot be put in the middle of a query plan. Generally, these are mechanism for accessing data and not designed for processing complex analytics workload.

Graph algorithm itself is a rich research area with many interesting problems and interesting algorithms. Graph database is getting more attentions and whether we need a special graph database is still an open problem. Google Pregel [7] popularized BSP processing [9] on graph data and many open source implementations [11] followed suit. Several researchers point out that scale up might be a better fit for graph algorithms than scale out. Here we make a note that transducer can easily gather data to one node. For example, to run SSSP we can gather data to one node and run Dijkstra algorithm instead of Bellman-Ford algorithm. There are other special purpose databases designed to handle a workload, for example, time series data [15].

Hadoop based system can implement map reduce paradigm using a procedural language like Java. In this ecosystem, Spark, which is probably ahead of others, can use DataFrame to combine SQL and other languages like Scala. DataFrame API [2] is still designed to process relational query, especially, allow user to implement specialized, more efficient relational operators. Spark also added GraphFrame [4] to process graph data. The underlying paradigm of GraphFrame is also BSP.  Unlike Pregel or GraphFrame that focus on graph API of nodes and edges, we decided to give user access to lower level BSP API.  Our treatment of the problem is a general framework and give user full control because the flexibility of tailoring an algorithm to one's own need is important.  For example, we know that Bill of Material query is really a traversal of a tree structure from root and we can simplify BSF algorithm and get better performance.   We are also considering releasing a core graph algorithm library of important algorithms like SSSP and triangle counting.

Many popular AI/Deep learning tools and libraries so far live in a parallel universe different from database.  Most of those libraries do no more than using a database connector to retrieve data.  Many commercial database systems are starting to integrate these tools.  We believe this is a very promising direction – ETL and data cleaning will be invaluable preprocessing tool for AI and in return, AI is an important tool in data integration.   Close interaction between database and AI requires deep integration – user generally prefer a unified programming interface/language to constantly switching between several systems.   An integrated system also saves operational and administrative cost.

Most commercial databases have extensions, or packages for machine learning.  In terms of brining deep analytics/algorithms into relation database, MADLib [12] is closest to our goal.  The approaches taken however are drastically different.  MADLib is probably the most impressive demonstration of currently available mechanisms, attacking problems with all kinds of weapons like UDF, UDT, cursor, temporary table, constructing and invoking SQL, etc. We encourage user to examine the SSSP implementation of MADLib versus our implementation. The complexity of the implementation has huge impact in practice, both on performance and day to day operation.  Debugging the correctness or analyzing the performance of a very complex implementation is a major task even for experts.

## 5   CONCLUSIONS AND FUTURE WORKS

We have implemented transducer in a parallel relational database system.  Transducer extends a relational database query engine with capability to run queries that many people consider not fit for relational query processing. As example, we have shown that time series query, graph query, and deep learning tools can be efficiently integrated within database. We plan to build a library of transducers, including but not limited to the most commonly used machine learning algorithm, graph algorithms, etc.

We want to explore more computing paradigms like BSP. We will investigate more efficient communication and synchronization mechanisms between transducers and other relational operators.  We will examine sharing transducer state between queries, or even between two Deepgreen clusters.

Another interesting direction is to exploit the hybrid computing environment.  Some algorithms can take advantage of hardware acceleration like GPU.   Others like some graph algorithm are memory intensive.  A hybrid environment, for example, adding a few machines with vast amount of memory into a big parallel database cluster, may have superior price performance ratio for many graph algorithms.